\documentclass[11pt,nofootinbib,floatfix,pra]{revtex4}%
\usepackage{amsfonts,amssymb,stmaryrd,latexsym,amsmath,braket}
\usepackage{graphicx,subfigure}
\usepackage[export]{adjustbox}
\usepackage{times}
\usepackage{bbm}
\usepackage{bm}
\usepackage{mathrsfs}
\usepackage{slashed}
\usepackage[version=3]{mhchem}
\usepackage{braket}
\usepackage{hyperref}
\usepackage{verbatim}
\usepackage[utf8]{inputenc}
\usepackage{tikz}
\usetikzlibrary{patterns}
\usepackage{graphicx,subfigure}

\usepackage{enumitem}

\usepackage{graphicx,subfigure}
\usepackage{times}
\usepackage{slashed}
\usepackage{braket}
\usepackage{verbatim}
\usepackage{multirow}
\usepackage[utf8]{inputenc}
\usepackage{array}
\usepackage{etoolbox}
\usepackage{mathtools}
\usepackage[percent]{overpic}
\newcommand{\PreserveBackslash}[1]{\let\temp=\\#1\let\\=\temp}
\newcolumntype{C}[1]{>{\PreserveBackslash\centering}p{#1}}
\newcolumntype{R}[1]{>{\PreserveBackslash\raggedleft}p{#1}}
\newcolumntype{L}[1]{>{\PreserveBackslash\raggedright}p{#1}}

\makeatletter
\let\orgdescriptionlabel\descriptionlabel
\renewcommand*{\descriptionlabel}[1]{%
  \let\orglabel\label
  \let\label\@gobble
  \phantomsection
  \edef\@currentlabel{#1}%
  \let\label\orglabel
  \orgdescriptionlabel{#1}%
}
\makeatother

\usepackage{color}

\usepackage{makecell}
\definecolor{DarkMidnightBlue}{rgb}{0.0, 0.04, 0.14}

\def\){\right)} 
\def\({\left(} 
\def\]{\right]} 
\def\[{\left[}

\hypersetup{hidelinks,
breaklinks=true,
colorlinks=true,
linkcolor=blue, 
citecolor=blue, 
urlcolor=DarkMidnightBlue,
bookmarksopen=false,    
pdftitle={Title},
pdfauthor={Author}}

\def\affONE{\affiliation{Department of Physics, Karamanoglu Mehmetbey University, Türkiye\\
and\\
School of Computing, Engineering, and Physical Sciences, University of the West of Scotland, Paisley, PA1 2BE, UK}}

\def\affTWO{\affiliation{School of Physics, Nankai University, Tianjin 300071, China}}

\def\affTHREE{\affiliation{Faculty of Engineering and Natural Sciences,\\ Gaziantep Islam Science and Technology University, Gaziantep 27010, Türkiye\\
and\\
King Fahd University of Petroleum and Minerals (KFUPM), 31261 Dhahran, Saudi Arabia}}

\begin{document}

\title{From~binding~and~saturation~to~criticality in~nuclear~matter with~lattice~effective~field~theory}

\author{Osman Agar}
\email{osmanagar@kmu.edu.tr}
\affONE

\author{Zhengxue Ren}
\email{zxren@nankai.edu.cn}
\affTWO

\author{Serdar Elhatisari}
\email{selhatisari@gmail.com}
\affTHREE

\begin{abstract}
We investigate the interaction dependence of the liquid-gas critical point of symmetric nuclear matter in finite-temperature lattice effective field theory. Building on the pinhole-trace algorithm, we benchmark a first-order perturbative treatment for representative Hamiltonian splittings and then compute the finite-temperature equation of state for a sequence of sign-friendly lattice Hamiltonians ranging from an SU(4)-symmetric interaction to Hamiltonians with physical ${}^{1}S_{0}$ and ${}^{3}S_{1}$ channel dependence and three improved leading-order Hamiltonians. The finite-temperature analysis is complemented by zero-temperature calculations of the symmetric-matter saturation point and the binding energies of selected nuclei within the same lattice framework. We find that the benchmarked perturbative strategy is quantitatively reliable in the thermodynamic regime studied. Across this Hamiltonian sequence, the LO Hamiltonians improve the overall description of finite-nucleus binding energies and move the zero-temperature saturation point toward the empirical region, while lowering the critical temperature from $15.33(6)$~MeV to $13.50(17)-13.71(19)$~MeV. These calculations show that finite-temperature criticality is not fixed by zero-temperature saturation and binding alone, and provide a complementary benchmark for future lattice interaction development.
\end{abstract}

\maketitle

\newpage

\section{Introduction
\label{sec:introduction}}

Understanding the equation of state (EoS) of nuclear matter away from zero temperature is a central problem in nuclear many-body theory. In isospin-symmetric matter, the liquid-gas phase transition and its critical point probe how the nuclear interaction shapes the free-energy surface at sub-saturation density. This regime is relevant to the interpretation of multifragmentation phenomena in intermediate-energy heavy-ion collisions and to the description of warm dilute matter in astrophysical environments. At the same time, empirical information on the critical region is indirect, since it is extracted from finite, dynamical, Coulomb-affected systems rather than from uniform infinite matter itself~\cite{PhysRevC.87.054622}.

Microscopic finite-temperature calculations of nuclear matter have advanced substantially over the last two decades. Chiral many-body perturbation theory and related thermodynamic studies have produced quantitative equations of state and liquid-gas critical parameters for homogeneous symmetric matter, while self-consistent Green’s-function (SCGF) calculations have clarified the dependence of the transition on both the nuclear Hamiltonian and the many-body approximation~\cite{Baldo:1999cvh,Soma:2009pf,Carbone:2013rca,Wellenhofer:2014hya,Carbone:2018kji,Carbone:2019pkr,Keller:2022crb}. At low densities, virial and quantum-statistical approaches have further emphasized the role of correlations and light clusters in warm dilute matter~\cite{Horowitz:2005nd,Horowitz:2005zv,Typel:2009sy}. Taken together, these studies show both the progress of the field and the continuing sensitivity of finite-temperature observables to the details of the interaction and its treatment~\cite{Soma:2009pf,Carbone:2013rca,Wellenhofer:2014hya,Carbone:2018kji,Carbone:2019pkr}.

Within chiral effective field theory (EFT), the thermodynamics of nuclear matter as well as the role of pion-exchange and many-body mechanisms in the EoS were discussed in Refs.~\cite{Holt:2013fwa,Holt:2014hma}. The impact of chiral two- and three-nucleon interactions on neutron and nuclear matter, including nuclear saturation and isospin-asymmetric matter, was further explored in Refs.~\cite{Hebeler:2009iv,Hebeler:2010xb,Drischler:2015eba,Drischler:2017wtt}. Ref.~\cite{Wellenhofer:2014hya} presented quantitative finite-temperature calculations of symmetric nuclear matter and its liquid-gas transition, while Ref.~\cite{Wellenhofer:2015qba} extended the analysis to isospin-asymmetric matter and showed the systematic shrinking of the coexistence region with increasing neutron excess. Thermal effects in the EoS and finite-temperature neutron matter were further examined in Refs.~\cite{Carbone:2019pkr,Keller:2020qhx}, and Ref.~\cite{Keller:2022crb} presented the EoS for arbitrary proton fraction and temperature, including preliminary microscopic ranges for the liquid-gas critical region of symmetric nuclear matter. A recent review, Ref.~\cite{Kaiser:2026msy}, summarizes this line of work and its implications for the liquid-gas transition in both symmetric and neutron-rich nuclear matter.

Lattice effective field theory provides a complementary route to this problem. Its main advantage is that nonperturbative correlations, including clustering and phase separation, arise dynamically in the Monte Carlo calculation rather than being imposed through a quasiparticle or cluster ansatz. Early lattice studies of nuclear thermodynamics were limited by the computational cost of finite-temperature simulations in large volumes~\cite{Muller:1999cp,Lee:2004si}. That bottleneck was overcome by the pinhole-trace algorithm, which made canonical \textit{ab initio} lattice thermodynamics feasible and enabled the first finite-temperature lattice determination, within this framework, of the symmetric-matter equation of state, liquid-vapor coexistence line, and critical point~\cite{Lu:2019nbg}. The same work also presented the first \textit{ab initio} lattice study of the temperature and density dependence of nuclear clustering~\cite{Lu:2019nbg}. More recently, finite-temperature nuclear lattice effective field theory (NLEFT) has also been applied to clustering in hot dilute nuclear matter using light-cluster distillation~\cite{Ren:2023ued}. These studies established finite-temperature lattice thermodynamics as a viable \textit{ab initio} program.

More broadly, recent years have seen substantial progress in NLEFT as a general \textit{ab initio} framework for nuclear structure, reactions, thermodynamics, and hypernuclear systems~\cite{Lee:2025req}. Beyond the early applications to light nuclei, clustering, scattering, and thermal matter, the development of high-fidelity lattice chiral interactions and wavefunction-matching techniques has enabled accurate descriptions of nucleon–nucleon phase shifts, binding energies, and charge radii over a broad mass range, including the saturation properties of symmetric nuclear matter and the neutron-matter EoS~\cite{Elhatisari:2022zrb}. The same framework has since been applied to structural and thermodynamic properties of hot neutron matter~\cite{Ma:2023ahg}, charge radii of silicon~\cite{Konig:2023rwe} and oxygen isotopes~\cite{Ren:2025vpe}, proton-rich systems such as 
${}^{22}$Si~\cite{Zhang:2024wfd}, the spectrum and geometric structure of beryllium isotopes~\cite{Shen:2024qzi}, pairing correlations in carbon and oxygen isotopes~\cite{Song:2025ofd}, and  the ground state energies of Sn isotopes near the proton dripline~\cite{Hildenbrand:2025voq}. NLEFT has also been extended to hypernuclei systems, hyper-neutron and hyper-nuclear matters~\cite{Hildenbrand:2024ypw,Tong:2024egi,Tong:2025sui,Tong:2025fzv}, and more recently to reaction and response observables through studies of neutron–alpha scattering and nuclear $\beta$ decay~\cite{Elhatisari:2024otn,Wang:2025swg,Elhatisari:2025fyu}. In parallel, ongoing methodological developments show that NLEFT is advancing not only in physics reach but also in algorithmic sophistication, including perturbative quantum Monte Carlo~\cite{Lu:2021tab}, partial-pinhole~\cite{Ren:2025vpe} and rank-one operator methods for radii~\cite{Ma:2023ahg}, multi-reference trial states~\cite{Wang:2025ahm} and dilated-coordinate methods for weakly bound systems~\cite{He:2025gir}.

In the earlier finite-temperature pinhole-trace study, the thermodynamic results were obtained for a single sign-friendly lattice interaction~\cite{Lu:2019nbg,Ren:2023ued}. The natural next nontrivial question is whether the liquid-gas critical region of symmetric nuclear matter evolves as the underlying lattice Hamiltonian is made more realistic, or whether the critical region probes different aspects of the interaction. Standard zero-temperature benchmarks mainly constrain the interaction through ground-state binding and saturation properties, together with a limited set of low-energy spin-isospin channels. The liquid-gas critical point, by contrast, is controlled by the free-energy surface of symmetric matter at finite temperature and sub-saturation density. Therefore, there is no a priori reason to expect that an interaction which performs better in the ground-state sector should also provide a better description of the finite-temperature critical region. Establishing whether such a correlation exists is important both for nuclear thermodynamics itself and for the broader program of constraining nuclear interactions with many-body observables.

In this work we address that problem within the lattice framework by combining sign-friendly Hamiltonians with a perturbative implementation of the pinhole-trace algorithm. We start from the SU(4)-symmetric interaction used in the nonperturbative part of our calculations, then introduce the ${}^{1}S_{0}$ and ${}^{3}S_{1}$ channel-dependent operators and three leading-order Hamiltonians with different short-range structures and different calibration outcomes. Within a common lattice framework, we analyze the finite-temperature EoS of symmetric nuclear matter together with selected zero-temperature observables in finite nuclei and symmetric nuclear matter. The objective is to determine how finite-temperature critical observables respond to different interaction design choices and how those responses compare with the corresponding changes in finite-nucleus binding energies and in the zero-temperature saturation point of symmetric nuclear matter.

The paper is organized as follows. Section~\ref{sec:Hamiltonian} defines the Hamiltonians used in this work. Section~\ref{sec:Benchmark} presents benchmark calculations for the perturbative pinhole-trace treatment. Section~\ref{sec:results-discussions} discusses the finite-temperature and zero-temperature results. Section~\ref{sec:summary} summarizes the main conclusions.

\section{Hamiltonian}
\label{sec:Hamiltonian}

In this work we study symmetric nuclear matter and selected finite nuclei on the lattice by using a class of sign-friendly Hamiltonians and their perturbative extensions. Our main goal is to understand how the thermodynamic observables, in particular the liquid-gas critical point, evolve as the interaction is taken from a simple SU(4)-symmetric form to realistic Hamiltonians.

We work on a periodic $L^{3}$ lattice with spatial lattice spacing $a=1.32~\mathrm{fm}$, which corresponds to a momentum cutoff $\Lambda=\pi/a \approx 471~\mathrm{MeV}$. Throughout this work, we use
units where $\hbar= c = k_{B} =1$. The Hamiltonian is written as
\begin{equation}
H = K + V ,
\end{equation}
where $K$ is the free nucleon Hamiltonian defined by using fast Fourier transforms to produce the exact dispersion relations and $V$ denotes the interaction part. Throughout this work we use the normal-ordered transfer matrix
\begin{equation}
M = : \exp(-a_{t} H) : ,
\end{equation}
with temporal lattice spacing $a_{t}$, where the colons denote normal ordering with respect to the nucleon creation and annihilation operators. Thermal observables are computed with the pinhole-trace formalism, while the zero-temperature energies are extracted from Euclidean-time extrapolations, and more details can be found in~\cite{Lee:2008fa,Lahde:2019npb,Lu:2019nbg}. To keep the notation transparent, in the following, we first define the smeared density operators that enter all interaction terms, then introduce the SU(4)-symmetric Hamiltonian and spin-isospin dependent $S$-wave operators, and finally specify the three LO Hamiltonians.

We start by defining the total nucleon density operator purely nonlocally smeared, 
\begin{equation}
\rho^{({\rm s)}}_{{\rm {NL}}}(\vec n)
=
\sum_{i,j = 0,1}
a^{({\rm s})\dagger}_{i,j}(\vec n)\,
a^{({\rm s})}_{i,j}(\vec n)\,,
\label{eq:rho-only-nonlocal}
\end{equation}
where $i$ is the spin index, $j$ is the isospin index, the nonlocally smeared annihilation and creation operators with the strength ${{\rm s}}$ are defined as,
\begin{align}
a^{({\rm s})}_{i,j}(\vec n)
= & 
a_{i,j}(\vec n)
+
{\rm s}
\sum_{|\vec n'|=1}
a_{i,j}(\vec n+\vec n'),
\label{eq:sNL-ann}\\
a^{({\rm s})\dagger}_{i,j}(\vec n)
= & 
a_{i,j}^{\dagger}(\vec n)
+
{\rm s}
\sum_{|\vec n'|=1}
a_{i,j}^{\dagger}(\vec n+\vec n')\,.
\label{eq:sNL-cre}
\end{align}
Next, we define the total nucleon density operator purely locally smeared, 
\begin{equation}
\rho^{({\rm s},d)}_{{\rm {L}}}(\vec n)
=
\sum_{i,j = 0,1}
a^{\dagger}_{i,j}(\vec n)\,
a^{\,}_{i,j}(\vec n)
+
{\rm s}
\sum_{1 \le |\vec n-\vec n'|^{2}\le d} \,\,
\sum_{i,j=0,1}
a^{\dagger}_{i,j}(\vec n')\,
a^{\,}_{i,j}(\vec n').
\label{eq:rho-only-local}
\end{equation}
where $d$ denotes the range and $s$ specifies the strength of the local smearing. Finally, we define both locally and nonlocally smeared total  nucleon density operator, 
\begin{equation}
\rho^{({\rm s},{\rm s'})}(\vec n)
=
\sum_{i,j}
{a}^{({\rm s'})\dagger}_{i,j}(\vec n)\,
{a}^{({\rm s'})}_{i,j}(\vec n)
+
{\rm s}
\sum_{|\vec n-\vec n'|=1}
\sum_{i,j}
{a}^{({\rm s'})\dagger}_{i,j}({\vec n}')\,
{a}^{({\rm s'})}_{i,j}({\vec n}').
\label{eq:rho-local-nlocal}
\end{equation}
In Eq.~(\ref{eq:rho-local-nlocal}), the first argument denotes the local smearing strength and the second the nonlocal smearing strength. We note that in Eqs.~(\ref{eq:sNL-ann}), (\ref{eq:sNL-cre}), and (\ref{eq:rho-local-nlocal}) the ranges of both local and nonlocal smearing are set to one lattice spacing, while in Eq.~(\ref{eq:rho-only-local}) the range of the local smearing, $d$, is a free parameter.

The starting point is the SU(4)-symmetric interaction~\cite{Lu:2018bat},
\begin{equation}
H_{\mathrm{SU(4)}} =
K
+
V^{\mathrm{SU(4)}}_{2N,{\rm s_{L}},{\rm s_{NL}}}
+
V^{\mathrm{SU(4)}}_{3N,{\rm s_{L}},{\rm s_{NL}}},
\label{eq:H_SU4_main}
\end{equation}
which is sign-friendly and can therefore be treated nonperturbatively with high precision. In compact form this interaction can be written as
\begin{equation}
H_{\mathrm{SU(4)}} =
K
\,\,
+
\,\,
\frac{C_{2}}{2}  \sum_{\vec n} 
: 
\left[
\rho^{({\rm s_{L}},{\rm s_{NL}})}(\vec n)
\right]^2
:
\,\, + \,\,
\frac{C_{3}}{6}  \sum_{\vec n} 
: 
\left[
\rho^{({\rm s_{L}},{\rm s_{NL}})}(\vec n)
\right]^3
:,
\label{eq:H_SU4_compact}
\end{equation}
The parameters ${\rm s_{L}}$ and ${\rm s_{NL}}$ control the strengths of local and nonlocal smearing of this Hamiltonian, respectively. Throughout this work, we use the parameter set given in Tab.~\ref{tab:LECs}.

\begin{table*}[t]
\centering
\caption{
Coupling constants and smearing parameters used in the Hamiltonians of this work.
The couplings $C_{{}^{1}S_{0}}$ and $C_{{}^{3}S_{1}}$ are given in lattice units. 
}
\label{tab:LECs}
\begin{tabular}{lcc}
\hline\hline
Interaction piece & Symbol & Value \\
\hline
SU(4) two-body coupling & $C_{2}$ & $-3.41\times 10^{-7}~\mathrm{MeV}^{-2}$ \\
SU(4) three-body coupling & $C_{3}$ & $-1.40\times 10^{-14}~\mathrm{MeV}^{-5}$ \\
Local smearing & ${\rm s_{L}}$ & $0.061$ \\
Nonlocal smearing & ${\rm s_{NL}}$ & $0.5$ \\\hline
${}^{1}S_{0}$ coupling & $C_{{}^{1}S_{0}}$ & $0.00338$ \\
${}^{3}S_{1}$ coupling & $C_{{}^{3}S_{1}}$ & $\hspace{-0.25cm}-0.00290$ \\
\hline\hline
\end{tabular}
\end{table*}

\begin{figure}[h!]
\includegraphics[width=0.7\textwidth]{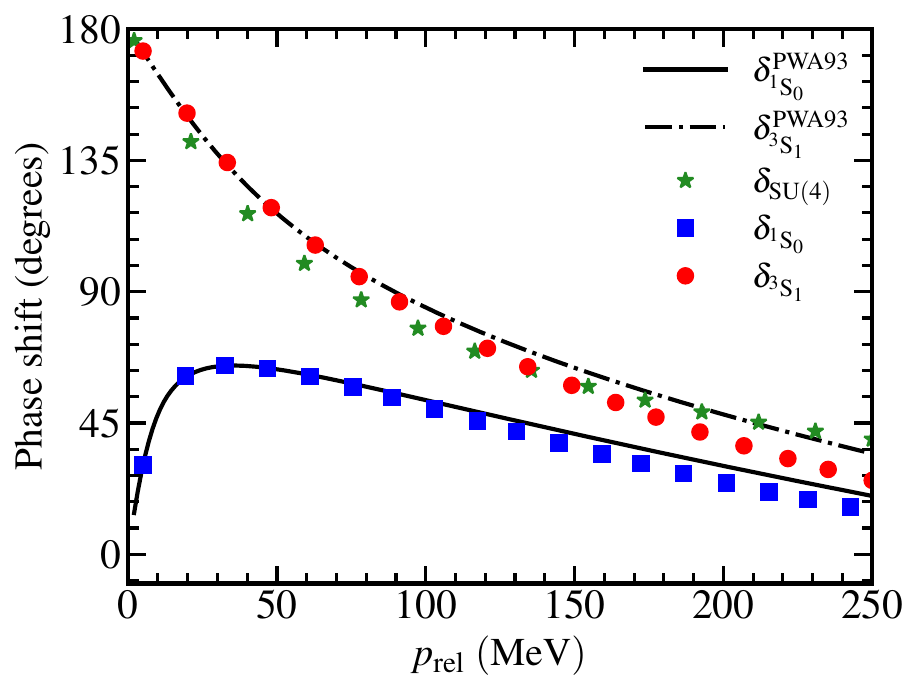}
\caption{
Lattice phase shifts in the physical ${}^{1}S_{0}$ and ${}^{3}S_{1}$ channels together with the corresponding SU(4)-symmetric interaction used in this work. The black solid and dashed lines denote the phase shift from the Nijmegen partial-wave analysis (PWA)~\cite{Stoks:1993tb}.
}
\label{fig:phase_shift}
\end{figure}

The next level of refinement is obtained by adding the two leading spin-isospin dependent $S$-wave operators,
\begin{equation}
H_{\mathrm{SU(4)}+S}
=
H_{\mathrm{SU(4)}}
+
V_{{}^{1}S_{0}}
+
V_{{}^{3}S_{1}} .
\label{eq:H_SU4_S}
\end{equation}
These terms break the artificial SU(4) symmetry in the relevant ${}^{1}S_{0}$ and ${}^{3}S_{1}$ channels. On the lattice, they are represented by the order-$Q^{0}$ operators
\begin{align}
V_{{}^{1}S_{0}}
&=
C_{{}^{1}S_{0}}
\sum_{\vec n}
V_{0,{}^{1}S_{0}}(\vec n),
\\
V_{{}^{3}S_{1}}
&=
C_{{}^{3}S_{1}}
\sum_{\vec n}
V_{0,{}^{3}S_{1}}(\vec n).
\end{align}
The coupling constants $C_{{}^{1}S_{0}}$ and $C_{{}^{3}S_{1}}$ are determined by fitting the lattice phase shifts to the phase shift from the Nijmegen partial-wave analysis (PWA)~\cite{Stoks:1993tb}, and the results are presented in Fig.~\ref{fig:phase_shift}, together with the phase shifts from the corresponding SU(4)-symmetric interaction used in this work. The determined values of these coupling constants are given in Tab.~\ref{tab:LECs}. In terms of the lattice pair operators, the interactions in ${}^{1}S_{0}$ and ${}^{3}S_{1}$ channels are
\begin{align}
V_{0,{}^{1}S_{0}}(\vec n)
&=
\sum_{I_{z}=-1,0,1}
\Bigl[
O^{0,{\rm s_{NL}}}_{0,0,0,0,1,I_{z}}(\vec n)
\Bigr]^{\dagger}
O^{0,{\rm s_{NL}}}_{0,0,0,0,1,I_{z}}(\vec n),
\\
V_{0,{}^{3}S_{1}}(\vec n)
&=
\sum_{J_{z}=-1,0,1}
\Bigl[
O^{0,{\rm s_{NL}}}_{1,0,1,J_{z},0,0}(\vec n)
\Bigr]^{\dagger}
O^{0,{\rm s_{NL}}}_{1,0,1,J_{z},0,0}(\vec n)\,.
\end{align}
These are purely nonlocally smeared and the smearing strength ${\rm s_{NL}}$ is given in Tab.~\ref{tab:LECs}. More details can be found in \cite{Li:2018ymw}.

Additionally, we construct three LO Hamiltonians in the framework of pionless EFT, denoted $H_{\rm LO_{1}}$, $H_{\rm LO_{2}}$ and $H_{\rm LO_{3}}$,
\begin{equation}
H_{\mathrm{LO}_i}
=
H_{\mathrm{SU(4)}}
+
V_{{}^{1}S_{0}}
+
V_{{}^{3}S_{1}} 
+
V_{3N}^{(i)},
\qquad i=1,2,3\,.
\label{eq:H_LOi}
\end{equation}
They include the physical ${}^{1}S_{0}$ and ${}^{3}S_{1}$ channel-dependent operators on top of the SU(4)-symmetric interaction, but they differ in the three-nucleon short-range SU(4)-symmetric corrections, which are constructed with different degrees of locality as, 
\begin{align}
V_{3N}^{(1)}
&=
V_{3N,{\rm L}}^{(1)}
+V_{3N,{\rm NL}}^{(1)}\,,
\\
V_{3N}^{(2)}
&=
V_{3N,{\rm L}}^{(2)}
+V_{3N,{\rm NL}}^{(2)},
\\
V_{3N}^{(3)}
&= 
V_{3N,{\rm L}}^{(3)}
+V_{3N,{\rm NL}}^{(3)} 
+
0.15\times
V^{\mathrm{SU(4)}}_{3N,{\rm s_{L}},{\rm s_{NL}}}
\,.
\end{align}
The explicit forms of these three-nucleon interactions are
\begin{align}
V_{3N,{\rm L}}^{(1)}
&=
\frac{0.03528}{6}
\sum_{\vec n}
:
\left[
\rho^{({\rm s_L},3)}_{{\rm L}}(\vec n)
\right]^{3}
:\,,
\\
V_{3N,{\rm L}}^{(2)}
&=
\frac{1}{6}
\sum_{\vec n}
:
\left\{
0.00658
\left[
\rho^{({\rm s_L},2)}_{{\rm L}}(\vec n)
\right]^{3}
+
0.03528
\left[
\rho^{({\rm s_L},3)}_{{\rm L}}(\vec n)
\right]^{3}
\right\}
: 
\,,
\\
V_{3N,{\rm L}}^{(3)}
&=
\frac{1}{6}
\sum_{\vec n}
:
\left\{
0.03294
\left[
\rho^{({\rm s_L},2)}_{{\rm L}}(\vec n)
\right]^{3}
+
0.03528
\left[
\rho^{({\rm s_L},3)}_{{\rm L}}(\vec n)
\right]^{3}
\right\}
: 
\,,
\end{align}
\begin{align}
V_{3N,{\rm NL}}^{(1)}
&= 
\frac{-0.00232}{6}
\sum_{\vec n}
:\,
\left\{
13.81
\left[
\rho_{\rm NL}^{(0.1)}(\vec n)
\right]^{3}
+
\, 3.34
\left[
\rho_{\rm NL}^{(0.2)}(\vec n)
\right]^{3}
+
\left[
\rho_{\rm NL}^{(0.3)}(\vec n)
\right]^{3}
\right\}
:
\,,
\\
V_{3N,{\rm NL}}^{(2)}
&= 
\frac{-0.00232}{6}
\sum_{\vec n}
:\,
\left\{
15.00
\left[
\rho_{\rm NL}^{(0.1)}(\vec n)
\right]^{3}
+
\, 3.34
\left[
\rho_{\rm NL}^{(0.2)}(\vec n)
\right]^{3}
+
\left[
\rho_{\rm NL}^{(0.3)}(\vec n)
\right]^{3}
\right\}
: \,,
\\
V_{3N,{\rm NL}}^{(3)}
&= 
\frac{-0.00232}{6}
\sum_{\vec n}
:\,
\left\{
15.00
\left[
\rho_{\rm NL}^{(0.1)}(\vec n)
\right]^{3}
+
\, 3.34
\left[
\rho_{\rm NL}^{(0.2)}(\vec n)
\right]^{3}
+
\left[
\rho_{\rm NL}^{(0.3)}(\vec n)
\right]^{3}
\right\}
: \,,
\end{align}
where the strength of the local smearing ${\rm s_{L}}$ is given in Tab.~\ref{tab:LECs}. The numerical coefficients entering these LO three-body interactions are determined using the optimization approach given in Ref.~\cite{Elhatisari:inprogress}.

It is important to comment on the meaning of the term LO pionless EFT in the present lattice formulation~\cite{Epelbaum:2008ga,Lee:2008fa}. The Hamiltonians introduced above are motivated by the leading $S$-wave structure of LO pionless EFT and by the associated low-energy spin-isospin channels. However, the actual lattice interactions contain local and nonlocal smearing, and the SU(4)-invariant three-body sector is also used to absorb lattice artifacts and to improve the many-body description. For this reason, it would be misleading to identify the systematic errors of the present calculation with those of a strict continuum LO pionless EFT calculation. In practice, the interactions used here should be viewed as sign-friendly lattice Hamiltonians that retain the leading low-energy $S$-wave physics while being optimized for \textit{ab initio} calculations of finite nuclei and nuclear matter.

In the next section we introduce the schematic perturbative splittings used to benchmark the accuracy of the first-order pinhole-trace treatment before turning to the physical results.

\section{Benchmark}
\label{sec:Benchmark}

Before presenting the physical finite-temperature results, we first perform a set of benchmark calculations in order to verify that the perturbative pinhole-trace formalism remains quantitatively under control in the density and temperature regime relevant for the present work. Also, at the end of this section we perform similar benchmark calculations for zero-temperature systems. The main purpose of this analysis is not to extract physical conclusions from the benchmark Hamiltonians themselves, but rather to test whether the first-order perturbative treatment accurately reproduces the results of the corresponding full Hamiltonian. This step is essential because all subsequent thermodynamic calculations rely on the same perturbative strategy.

\begin{table}[h!]
\centering
\caption{Forms of Hamiltonians used in the benchmark calculations.}
\begin{tabular}{c|c|c}
\hline\hline
 \hspace{1cm} & $H^{(0)}_{i}$ (Non-perturbative) & $H^{(1)}_{i}$ (Perturbative) \\ \hline
 $H$ & $V_{2N,{\rm s_{L}},{\rm s_{NL}}}^{\rm SU(4)}
+ V_{3N,{\rm s_{L}},{\rm s_{NL}}}^{\rm SU(4)}$ & 0  \\ \hline
 $H_{1}$    & $0.9\times V_{2N,{\rm s_{L}},{\rm s_{NL}}}^{\rm SU(4)}
+ V_{3N,{\rm s_{L}},{\rm s_{NL}}}^{\rm SU(4)}$ & $0.1\times V_{2N,{\rm s_{L}},{\rm s_{NL}}}^{\rm SU(4)}$  \\  \hline
 $H_{2}$   & $0.9\times V_{2N,{\rm s_{L}},{\rm s_{NL}}}^{\rm SU(4)}
+ 0.8\times V_{3N,{\rm s_{L}},{\rm s_{NL}}}^{\rm SU(4)}$ & $0.1\times V_{2N,{\rm s_{L}},{\rm s_{NL}}}^{\rm SU(4)}
+ 0.2\times V_{3N,{\rm s_{L}},{\rm s_{NL}}}^{\rm SU(4)}$  \\\hline
\hline
\end{tabular}
\label{tab:benchmark-hamiltonians}
\end{table}

To this end, we consider the family of Hamiltonian decompositions listed in Table~\ref{tab:benchmark-hamiltonians}. In each case, the Hamiltonian is written as
\begin{equation}
H_i = H_i^{(0)} + H_i^{(1)},
\end{equation}
where $H_i^{(0)}$ is treated nonperturbatively in the pinhole-trace calculation, while $H_i^{(1)}$ is included as a first-order perturbative correction. The benchmark set is designed to probe different ways of distributing the two-body and three-body interaction strengths between the nonperturbative and perturbative sectors. In this way, one can test whether the perturbative treatment remains accurate not only for a single special choice of splitting, but over a representative range of decompositions.

In the canonical formalism, the chemical potential is extracted from particle insertion and removal observables. Following the Widom-insertion strategy~\cite{Widom_1963} adapted to the pinhole-trace algorithm~\cite{Lu:2019nbg}, the chemical potential is obtained from
\begin{equation}
\mu
=
\frac{T}{2}
\ln\!\left[
A(A+1)\,
\frac{\langle \mathcal{B}_{-1}\rangle}
{\langle \mathcal{B}_{1}\rangle}
\right],
\label{eq:mu_benchmark}
\end{equation}
where $A$ is the nucleon number, $T$ is the temperature, and $\mathcal{B}_{1}$ and $\mathcal{B}_{-1}$ denote the insertion and removal estimators, respectively. We denote the single-particle quantum numbers by
$c_i=(\vec n_i,\sigma_i,\tau_i)$, where $\vec n_i$ is the lattice coordinate, $\sigma_i$ is the spin, and $\tau_i$ is the isospin.
We use the abbreviations $\vec c=\{c_1,\ldots,c_A\}$ for the many-body configuration and
$\vec s=\{s_1,\ldots,s_{L_t}\}$ for the auxiliary-field configuration, where $L_t$ is the number of temporal lattice steps.
The quantity $P(\vec s,\vec c)$ denotes the corresponding Monte Carlo sampling weight. For the purely nonperturbative calculation we define
\begin{align}
\mathcal{B}_{1}^{(0)}
= &
\sum_{c'}
\frac{\langle \vec c \cup c'|M(s_{L_t})\cdots M(s_1)|\vec c \cup c'\rangle}
{P(\vec s,\vec c)},
\nonumber\\
\mathcal{B}_{-1}^{(0)}
= &
\sum_{c_i}
\frac{\langle \vec c\setminus c_i|M(s_{L_t})\cdots M(s_1)|\vec c\setminus c_i\rangle}
{P(\vec s,\vec c)},
\label{eq:Bpm0}
\end{align}
while the first-order contributions are written as
\begin{align}
\mathcal{B}_{1}^{(1)}
= & L_{t}
\sum_{c'}
\frac{\langle \vec c \cup c'|M^{(1)}\,M(s_{L_t-1})\cdots M(s_1)|\vec c \cup c'\rangle}
{P(\vec s,\vec c)},
\nonumber\\
\mathcal{B}_{-1}^{(1)}
= &  L_{t}
\sum_{c_i}
\frac{\langle \vec c\setminus c_i|M^{(1)}\,M(s_{L_t-1})\cdots M(s_1)|\vec c\setminus c_i\rangle}
{P(\vec s,\vec c)}\,.
\label{eq:Bpm1}
\end{align}
Here the sum over $c'$ runs over the single-particle quantum numbers of the inserted nucleon, while $c_i$ denotes one of the nucleons already present in the configuration $\vec c$. Similarly, for the normalization factor we write
\begin{align}
\mathcal{M}^{(0)}
=
\frac{\braket{\vec{c}
|M(s_{L_{t})}
\ldots
M({s_{1})}
|\vec{c}}}
{P(\vec s,\vec c)},
\qquad
\mathcal{M}^{(1)}
= L_{t}
\frac{
\braket{\vec{c}
|M^{(1)}M(s_{L_{t}-1)}
\ldots
M({s_{1})}
|\vec{c}}}
{P(\vec s,\vec c)}.
\label{eq:M01}
\end{align}
Expanding to first order in the perturbation gives
\begin{align}
\frac{\mathcal{B}_{\pm1}}{\mathcal{M}}
=
\frac
{\mathcal{B}^{(0)}_{\pm1} + \mathcal{B}^{(1)}_{\pm1}}
{\mathcal{M}^{(0)} + \mathcal{M}^{(1)}}
=
\frac
{\mathcal{B}^{(0)}_{\pm1} }
{\mathcal{M}^{(0)}}
+
\frac
{\mathcal{B}^{(1)}_{\pm1}}
{\mathcal{M}^{(0)}}
-
\frac
{\mathcal{B}^{(0)}_{\pm1}    \mathcal{M}^{(1)}}
{\mathcal{M}^{(0)} \mathcal{M}^{(0)}}
+
\mathcal{O}\!\left((M^{(1)})^2\right).
\label{eq:first_order_ratio}
\end{align}

\begin{figure}[h!]
\includegraphics[width=0.48\textwidth]{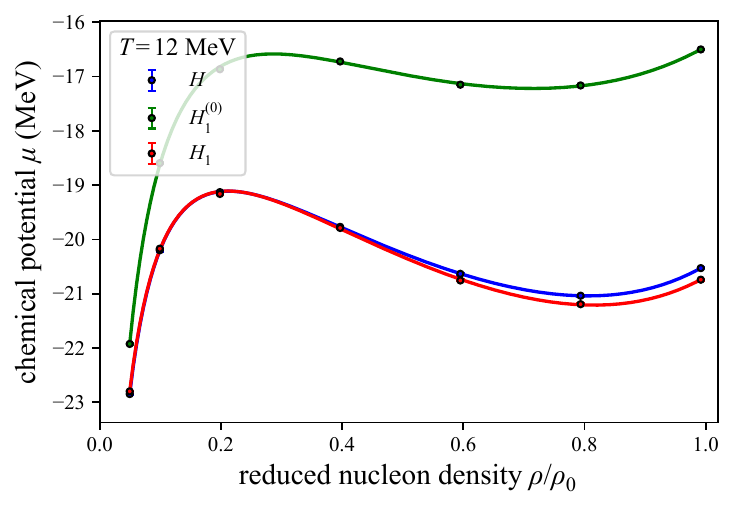}
\includegraphics[width=0.48\textwidth]{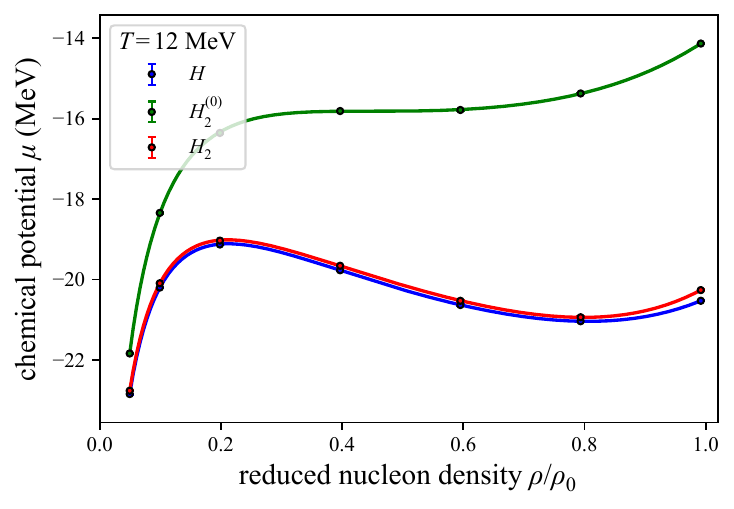}
\caption{
Benchmark results for the perturbative pinhole-trace calculations at $T=12$ MeV.
The chemical potential $\mu$ is shown as a function of the reduced nucleon density
$\rho/\rho_{0}$ for the Hamiltonian splittings $H_{1}$ (left) and $H_{2}$ (right),
defined in Table~\ref{tab:benchmark-hamiltonians}, where $\rho_{0}=0.164$~fm${}^{-3}$.
In each panel, the blue points and curve denote the result obtained with the full
Hamiltonian $H$, the green points and curve show the result obtained with the
nonperturbative reference Hamiltonian $H_{i}^{(0)}$, and the red points and curve
correspond to the first-order corrected result
$H_{i}=H_{i}^{(0)}+H_{i}^{(1)}$.}
\label{fig:benchmark-T12}
\end{figure}

\begin{figure}[h!]
\includegraphics[width=0.48\textwidth]{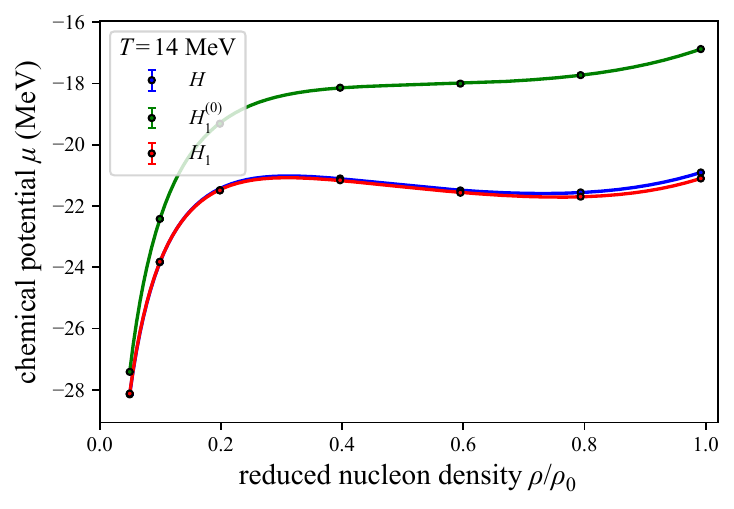}
\includegraphics[width=0.48\textwidth]{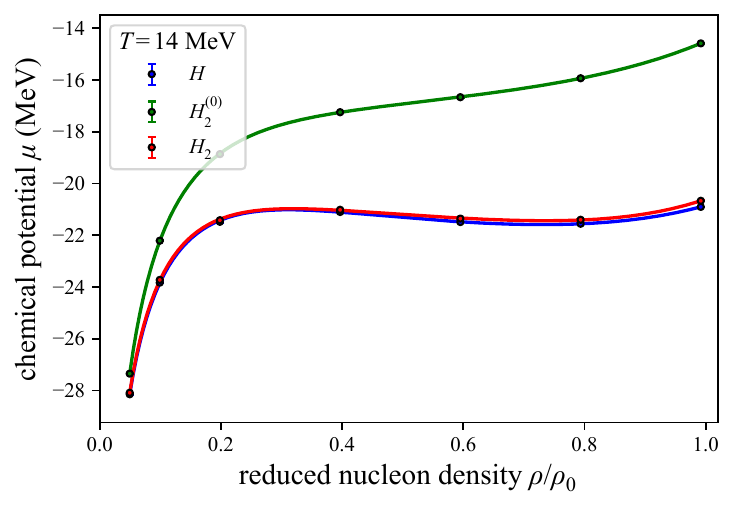}
\caption{Same as in Fig.~\ref{fig:benchmark-T12}, but at $T=14$ MeV.}
\label{fig:benchmark-T14}
\end{figure}

\begin{figure}[h!]
\includegraphics[width=0.48\textwidth]{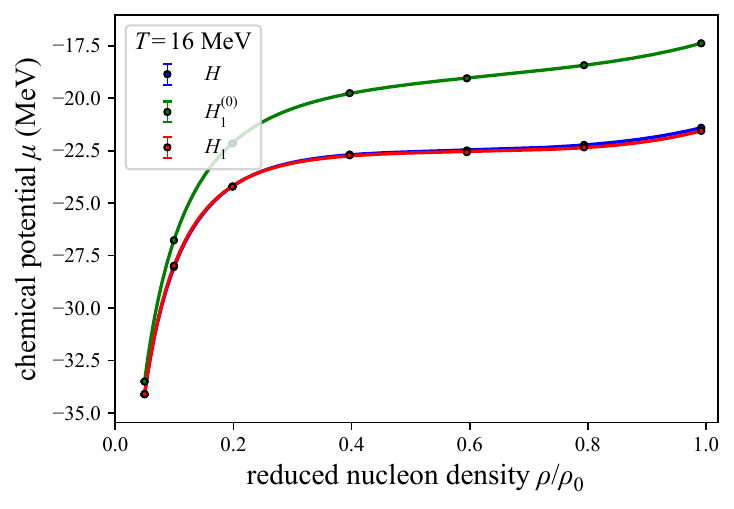}
\includegraphics[width=0.48\textwidth]{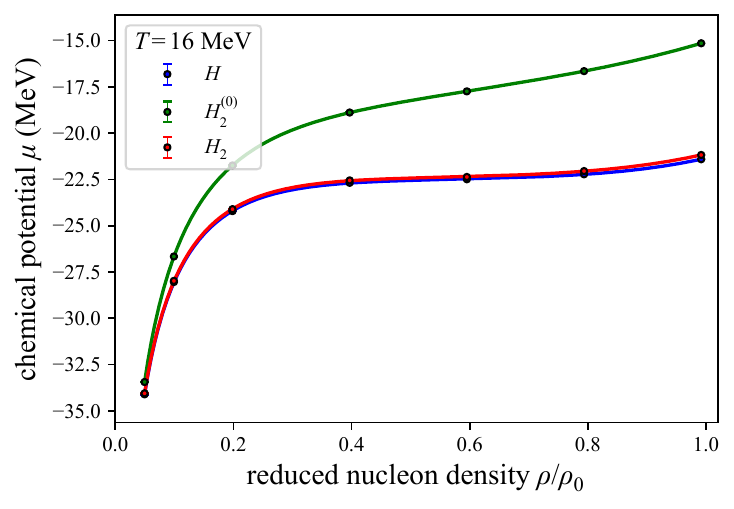}
\caption{Same as in Fig.~\ref{fig:benchmark-T12}, but at $T=16$ MeV.}
\label{fig:benchmark-T16}
\end{figure}

The benchmark results for $T=12$, $14$, and $16$ MeV are shown in Figs.~\ref{fig:benchmark-T12}--\ref{fig:benchmark-T16}. In each panel we compare three quantities, which are the result obtained with the full Hamiltonian $H$, the result obtained with the nonperturbative part $H_i^{(0)}$ only, and the result obtained after adding the first-order perturbative correction, denoted by $H_i$. The main observation is that the nonperturbative reference Hamiltonian $H_i^{(0)}$ can differ significantly from the full result, as expected from the omitted interaction terms. However, once the first-order contribution is included, the corresponding $H_i$ curve closely follows the result of the full Hamiltonian over the entire density range considered.

This pattern is precisely what is required for a controlled perturbative treatment. The perturbative correction is not artificially small since it produces a visible and physically meaningful shift relative to $H_i^{(0)}$, showing that the omitted terms are indeed important. At the same time, the first-order corrected result remains very close to the full calculation, which demonstrates that the perturbative expansion is convergent and quantitatively reliable in the present thermodynamic regime. In other words, the benchmark confirms that the chosen nonperturbative reference Hamiltonians are sufficiently close to the full interaction to serve as stable starting points for the perturbative pinhole-trace calculation.

\begin{table*}[!htb]
\centering
\caption{Benchmark results for zero-temperature symmetric nuclear matter near saturation. For each benchmark system, the table compares the full Hamiltonian result $\langle H \rangle$, the nonperturbative reference $\langle H_i^{(0)} \rangle$, and the first-order corrected result $\langle H_i \rangle$. The corrected energies remain very close to the full-Hamiltonian values at densities $\rho=0.163$--$0.169~\mathrm{fm}^{-3}$, showing that the perturbative treatment is quantitatively reliable also in the zero-temperature saturation region. All values are taken at Euclidean time $\tau = 0.4~\mathrm{MeV}^{-1}$. As shown by the Euclidean-time extrapolations for symmetric nuclear matter, the energies per nucleon are already close to their asymptotic plateau by this Euclidean time, see Fig.~\ref{fig:appendix-snm-euclid}.}
\label{tab:benchmark-SNM}
\begin{tabular}{lcc}
\hline\hline
Hamiltonian & \hspace{0.5cm}$E$ (MeV)\hspace{0.5cm} & \hspace{0.5cm}$E/A$ (MeV)\hspace{0.5cm} \\
\hline
\multicolumn{3}{c}{$A=48$, $L=6.6~\mathrm{fm}$, $\rho=0.169~\mathrm{fm}^{-3}$} \\
\hline
$\braket{H}$ & -781.83(85) & -16.288(18) \\
\hline
$\braket{H_{1}^{(0)}}$ & -613.91(59) & -12.790(12) \\
$\braket{H_{1}}$       & -780.48(61) & -16.260(13) \\
\hline
$\braket{H_{2}^{(0)}}$ & -557.41(59) & -11.613(12) \\
$\braket{H_{2}}$       & -779.02(61) & -16.230(13) \\
\hline
\multicolumn{3}{c}{$A=80$, $L=7.9~\mathrm{fm}$, $\rho=0.163~\mathrm{fm}^{-3}$} \\
\hline
$\braket{H}$ & -1300.4(12) & -16.255(15) \\
\hline
$\braket{H_{1}^{(0)}}$ & -1022.8(11) & -12.785(14) \\
$\braket{H_{1}}$       & -1295.7(11) & -16.196(14) \\
\hline
$\braket{H_{2}^{(0)}}$ & -930.67(102) & -11.633(13) \\
$\braket{H_{2}}$       & -1290.7(10) & -16.134(13) \\
\hline\hline
\end{tabular}
\end{table*}
Since the benchmark analysis above was formulated for finite-temperature observables, it is useful to verify that the same perturbative strategy also remains accurate for zero-temperature symmetric nuclear matter in the vicinity of the saturation region. Table~\ref{tab:benchmark-SNM} shows such a test for two benchmark systems with densities $\rho=0.169~\mathrm{fm}^{-3}$ and $\rho=0.163~\mathrm{fm}^{-3}$, both close to the empirical saturation regime. All values in the table are evaluated at Euclidean time $\tau = a_{t} \times L_{t} = 0.4~\mathrm{MeV}^{-1}$. As seen from the Euclidean-time extrapolations of the symmetric nuclear matter energies per nucleon given in Fig.~\ref{fig:appendix-snm-euclid} in the Appendix, the results have already reached an essentially converged plateau by this projection time, so no significant residual Euclidean-time contamination is expected at the level relevant for the present benchmark.  As in the finite-temperature benchmark, the nonperturbative reference Hamiltonians $H_i^{(0)}$ differ substantially from the full result, whereas the inclusion of the first-order perturbative correction brings the energies back into very good agreement with the full Hamiltonian. For $A=48$, the residual differences between $\langle H_i \rangle$ and $\langle H \rangle$ are only $1.35$ and $2.81$ MeV, while for $A=80$ they are $4.7$ and $9.7$ MeV, corresponding to about $0.03$--$0.12$ MeV per nucleon.

Overall, the benchmark calculations provide strong evidence that the perturbative pinhole-trace formalism is controlled for the class of Hamiltonians considered here. This validation step justifies the use of the same framework in the finite-temperature and zero-temperature calculations presented in the following sections.

\section{Results and Discussion\label{sec:results-discussions}}

\begin{table*}[t]
\centering
\caption{
Critical-point and saturation properties obtained for the Hamiltonians considered in this work.
Shown are the critical temperature $T_c$, critical pressure $P_c$, critical density $\rho_c$,
saturation density $\rho_{\mathrm{sat}}$, and saturation energy per nucleon
$E_{\mathrm{sat}}/A$.
The quoted uncertainties for the theoretical results are the bootstrap errors.
Phenomenological empirical estimates are shown in the last column for the critical-point observables. The saturation point comparison is shown separately in Fig.~\ref{fig:snm-eos-zeroT}.}
\label{tab:critical_saturation_all}
\begin{tabular}{lcccccc}
\hline\hline
 & $H_{\mathrm{SU(4)}}$
 & $H_{\mathrm{SU(4)}+S}$
 & $H_{\mathrm{LO}_1}$
 & $H_{\mathrm{LO}_2}$
 & $H_{\mathrm{LO}_3}$
 & Exp.~\cite{Elliott:2013pna} \\
\hline
$T_c$ (MeV)
& $15.33(6)$
& $15.13(18)$
& $13.71(19)$
& $13.54(18)$
& $13.50(17)$
& $17.9(4)$ \\

$P_c$ (MeV/fm$^3$)
& $0.274(4)$
& $0.279(16)$
& $0.181(11)$
& $0.172(10)$
& $0.166(9)$
& $0.31(7)$ \\

$\rho_c$ (fm$^{-3}$)
& $0.0884(30)$
& $0.0983(64)$
& $0.0720(46)$
& $0.0689(35)$
& $0.0657(26)$
& $0.06(1)$ \\

$\rho_{\mathrm{sat}}$ (fm$^{-3}$)
& $0.2188(2)$
& $0.2177(2)$
& $0.1818(2)$
& $0.1783(2)$
& $0.1717(5)$
& --- \\

$E_{\mathrm{sat}}/A$ (MeV)
& $-17.0930(41)$
& $-16.4885(46)$
& $-15.9253(56)$
& $-15.7592(84)$
& $-15.8317(261)$
& --- \\
\hline\hline
\end{tabular}
\end{table*}

\begin{table}[h!]
\centering
\caption{Comparison of the calculated ground-state binding energies, in MeV, against experimental values. The $\braket{H_{\rm LO_{1}}}$, $\braket{H_{\rm LO_{2}}}$, and $\braket{H_{\rm LO_{3}}}$ columns are the LO results.}
\label{tab:binding_energies_samples}
\begin{tabular}{lcccccc}
\hline\hline
Nucleus & $\braket{H_{\rm SU(4)}}$ & $\braket{H_{{\rm SU(4)}+S}}$ & $\braket{H_{\rm LO_{1}}}$  & $\braket{H_{\rm LO_{2}}}$  & $\braket{H_{\rm LO_{3}}}$  & $E_{\text{Exp}}$ \\
\hline
$^{3}\mathrm{H}$   & $-8.396(69) $ & $-7.773(69) $ & $-8.330(15)  $ & $-8.335(15)  $ & $-8.472(16)  $ & $-8.48 $ \\
$^{4}\mathrm{He}$  & $-28.315(11)$ & $-26.864(11)$ & $-30.1280(30)$ & $-30.1512(26)$ & $-30.8375(47)$ & $-28.30 $ \\
$^{6}\mathrm{He}$  & $-30.31(24) $ & $-26.34(25) $ & $-29.636(56) $ & $-29.653(55) $ & $-30.372(56) $ & $-29.27 $ \\
$^{6}\mathrm{Li}$  & $-29.39(43) $ & $-29.90(43) $ & $-33.206(97) $ & $-33.225(97) $ & $-33.942(98) $ & $-31.99 $ \\
$^{8}\mathrm{Be}$  & $-56.27(12) $ & $-53.30(12) $ & $-59.878(27) $ & $-59.917(26) $ & $-61.318(28) $ & $-56.50 $ \\
$^{10}\mathrm{B}$  & $-57.3(14)  $ & $-55.9(14)  $ & $-62.42(32)  $ & $-62.44(32)  $ & $-63.89(32)  $ & $-64.75 $ \\
$^{12}\mathrm{C}$  & $-84.34(15) $ & $-79.83(15) $ & $-89.538(33) $ & $-89.566(33) $ & $-91.697(35) $ & $-92.16 $ \\
$^{16}\mathrm{O}$  & $-119.06(41)$ & $-112.65(41)$ & $-124.144(93)$ & $-123.886(92)$ & $-126.761(95)$ & $-127.62$ \\
$^{20}\mathrm{Ne}$ & $-149.85(11)$ & $-141.85(11)$ & $-155.955(26)$ & $-155.622(25)$ & $-159.189(35)$ & $-160.65$ \\
$^{24}\mathrm{Mg}$ & $-180.84(18)$ & $-171.11(18)$ & $-188.034(43)$ & $-187.594(42)$ & $-191.917(52)$ & $-198.26$ \\
$^{28}\mathrm{Si}$ & $-213.57(30)$ & $-202.05(30)$ & $-221.132(69)$ & $-220.474(68)$ & $-225.503(79)$ & $-236.54$ \\
$^{32}\mathrm{S}$  & $-248.55(51)$ & $-235.33(51)$ & $-256.11(12) $ & $-255.15(12) $ & $-260.81(13) $ & $-271.78$ \\
$^{36}\mathrm{Ar}$ & $-285.67(29)$ & $-270.18(29)$ & $-291.952(67)$ & $-290.590(67)$ & $-296.769(93)$ & $-306.72$ \\
$^{40}\mathrm{Ca}$ & $-328.89(22)$ & $-311.30(23)$ & $-332.758(55)$ & $-330.823(57)$ & $-337.29(10) $ & $-342.05$ \\
\hline\hline
\end{tabular}
\end{table}
\begin{figure}[htb!]
    \centering  \includegraphics[width=0.7\textwidth]{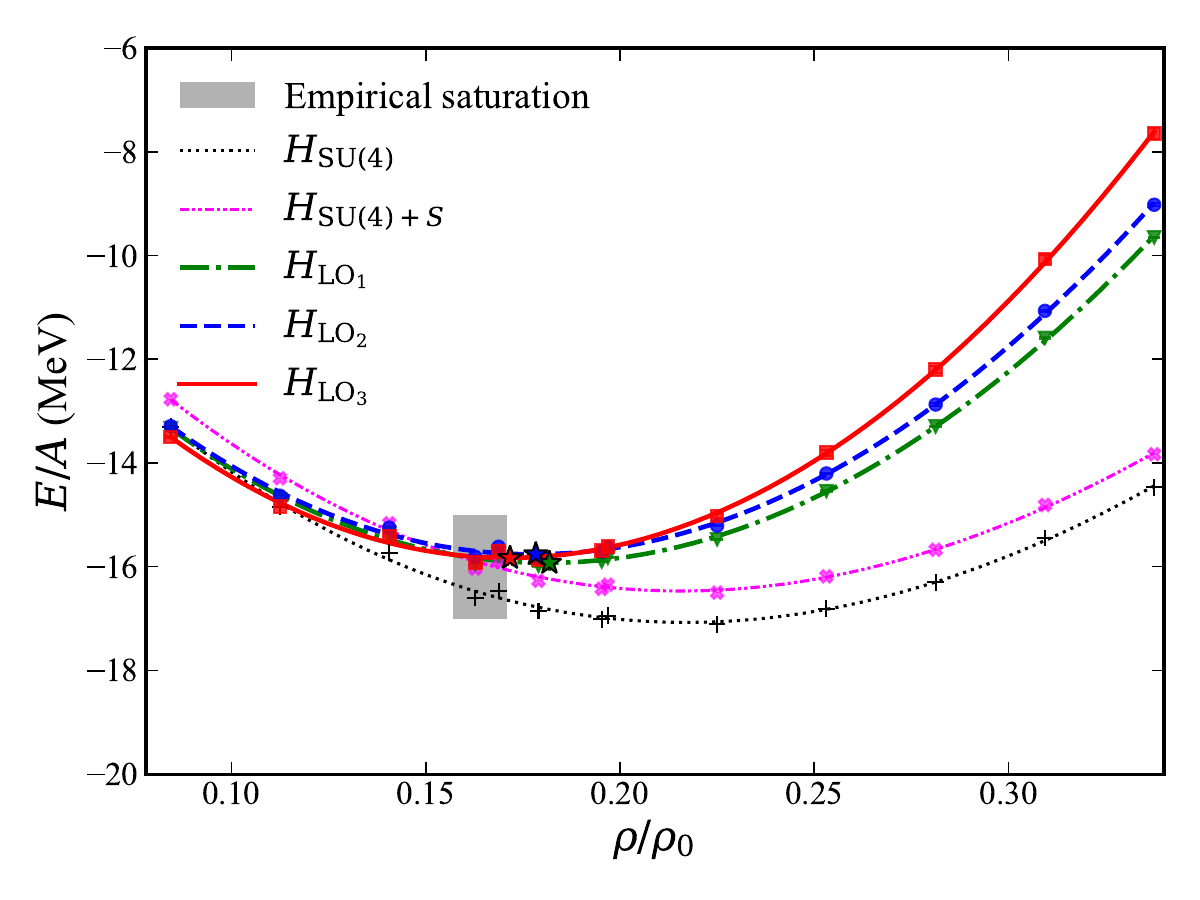}
\caption{Zero-temperature symmetric nuclear matter equation of state for the Hamiltonians considered in this work. The energy per nucleon $E/A$ is plotted as a function of the reduced density $\rho/\rho_{0}$. The dotted black, dashed blue, and solid red curves correspond to $H_{\mathrm{SU(4)}}$, $H_{\mathrm{SU(4)}+S}$, and the ``LO pionless-EFT'' Hamiltonians, respectively. The gray box shows the empirical saturation region ($0.16\pm0.01$~fm${}^{-3}$, $-16.0\pm1.0$~MeV)~\cite{Bethe:1971xm}}.
\label{fig:snm-eos-zeroT}
\end{figure}

Table~\ref{tab:critical_saturation_all} summarizes the critical-point and saturation properties obtained for all Hamiltonians considered in this work. The SU(4)-symmetric Hamiltonian gives the largest critical temperature, $T_{c}=15.33(6)$ MeV. The inclusion of the physical ${}^{1}S_{0}$ and ${}^{3}S_{1}$ interactions lowers it slightly to $15.13(18)$ MeV. For the three ``LO pionless-EFT'' Hamiltonians we find $T_{c}=13.71(19)$, $13.54(18)$, and $13.50(17)$ MeV. A similar trend is seen for the critical pressure and the critical density. At zero temperature, by contrast, the saturation point moves toward the empirical region as the critical temperature decreases. The SU(4)-symmetric Hamiltonian gives $\rho_{\mathrm{sat}}=0.2188(2)~\mathrm{fm}^{-3}$ and $E_{\mathrm{sat}}/A=-17.0930(41)$ MeV, while the ``LO pionless-EFT'' Hamiltonians give $\rho_{\mathrm{sat}}=0.1818(2)$, $0.1783(2)$, and $0.1717(5)~\mathrm{fm}^{-3}$, and $E_{\mathrm{sat}}/A=-15.9253(56)$, $-15.7592(84)$, and $-15.8317(261)$~MeV. The empirical critical-point values quoted in the last column should be interpreted as phenomenological reference estimates, since they are inferred indirectly from finite, dynamical, Coulomb-affected systems rather than from uniform infinite matter itself.

Table~\ref{tab:binding_energies_samples} and Fig.~\ref{fig:snm-eos-zeroT} show how the same Hamiltonian sequence behaves at zero temperature in finite nuclei and in symmetric nuclear matter. The SU(4)-symmetric Hamiltonian gives a reasonable overall description of the lighter nuclei, but the nuclei become systematically less bound as the mass number increases. The inclusion of the ${}^{1}S_{0}$ and ${}^{3}S_{1}$ interactions alone does not remove this trend. The three ``LO pionless-EFT'' Hamiltonians lead to a clear improvement and give a more consistent description of the binding energies over the full mass range considered here. Figure~\ref{fig:snm-eos-zeroT} shows the same pattern. The ``LO pionless-EFT'' Hamiltonians move the saturation point toward the empirical region, whereas $H_{\mathrm{SU(4)}}$ and $H_{\mathrm{SU(4)}+S}$ remain more deeply bound and saturate at larger densities. The spread among the three LO samples remains small compared with the overall shift from $H_{\mathrm{SU(4)}}$ and $H_{\mathrm{SU(4)}+S}$ to the LO Hamiltonians.

A first indication of the interaction dependence is shown in Fig.~\ref{fig:phase_shift}. The SU(4)-symmetric interaction produces a noticeably stronger attraction at larger relative momenta than the lattice phase shifts in the physical ${}^{1}S_{0}$ and ${}^{3}S_{1}$ channels. Once the interaction is resolved into the $S$-wave channels, the phase shifts move closer to the empirical behavior and the excessive attraction at higher momenta is reduced. This provides a useful qualitative reference point for both the zero-temperature improvements seen in Table~\ref{tab:binding_energies_samples} and Fig.~\ref{fig:snm-eos-zeroT}, and the finite-temperature trends discussed below.

For each Hamiltonian, the finite-temperature chemical-potential isotherms are analyzed by fitting the lattice results at fixed temperature with a fifth-order virial-like expansion,
\begin{equation}
\mu(\rho,T)
=
a_{0}(T)+a_{1}(T)\ln\rho+
\sum_{i = 1}^{4} a_{i+1}(T) \,\, \rho^{i}.
\label{eq:mu_virial_fit}
\end{equation}
The coefficients $a_{i}(T)$ are determined separately for each isotherm by a weighted least-squares fit using the Monte Carlo uncertainties of the lattice data. For temperatures between the simulated values, each coefficient is then interpolated by cubic splines as a function of $T$. In this way we obtain a smooth representation of $\mu(\rho,T)$ over the full density and temperature range considered in this work. Eq.~(\ref{eq:mu_virial_fit}) is used here only as a convenient parametrization of the lattice isotherms, and the fitted coefficients are not interpreted as physical virial coefficients. Using the fitted chemical potential, we construct the free-energy density and pressure analytically and determine the liquid-vapor coexistence line from the common-tangent Maxwell construction. The quoted uncertainties of $T_{c}$, $P_{c}$, and $\rho_{c}$ are extracted by bootstrap resampling of the lattice results and therefore reflect statistical propagation of the Monte Carlo errors. They do not include systematic effects associated with the fit ansatz, spline interpolation, Maxwell construction, or EFT truncation.

\begin{figure}[htb!]
    \centering  \includegraphics[width=0.55\textwidth]{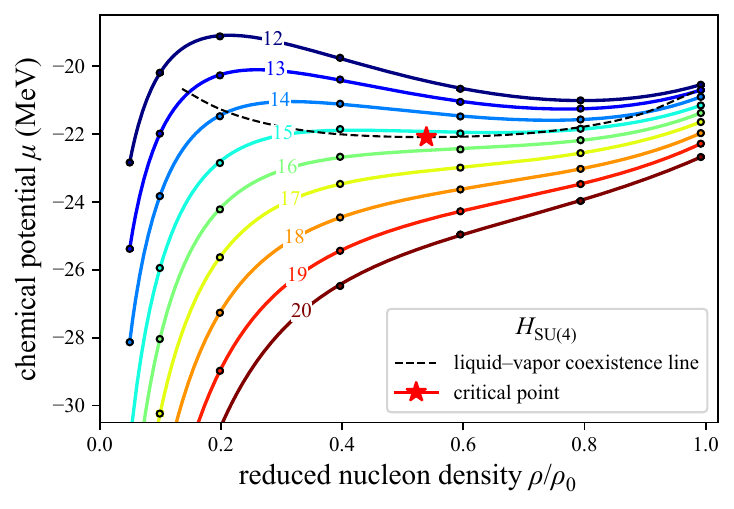}
\caption{Chemical potential $\mu$ as a function of the reduced nucleon density $\rho/\rho_{0}$ for the SU(4)-symmetric Hamiltonian $H_{\mathrm{SU(4)}}$, where $\rho_{0}=0.164$~fm${}^{-3}$. The symbols show the lattice results at different temperatures, while the solid curves are obtained from weighted fits of the individual isotherms using the fifth-order virial-like expansion of Eq.~(\ref{eq:mu_virial_fit}). The dashed curve denotes the liquid-vapor coexistence line extracted from the Maxwell construction, and the star marks the calculated critical point.}
\label{fig:mu-vs-rho-Hsu4}
\end{figure}
\begin{figure}[htb!]
    \centering  \includegraphics[width=0.55\textwidth]{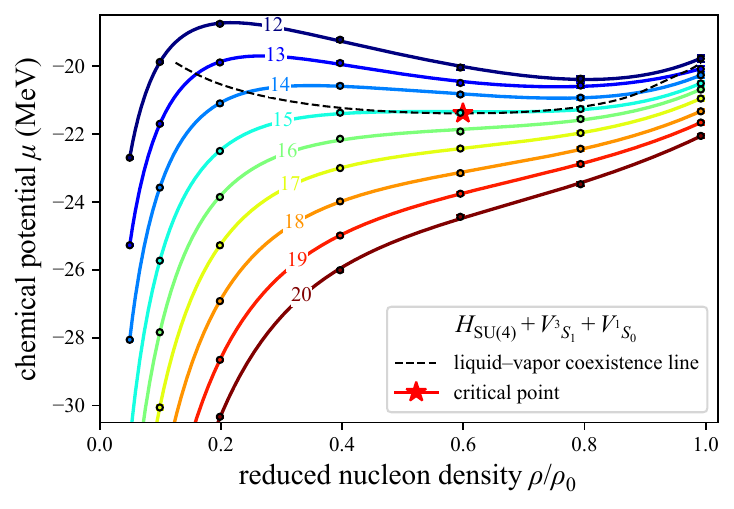}
\caption{Same as in Fig.~\ref{fig:mu-vs-rho-Hsu4}, but for $H_{\mathrm{SU(4)}+S} = H_{\mathrm{SU(4)}} + V_{{}^{1}S_{0}} + V_{{}^{3}S_{1}}$.}
\label{fig:mu-vs-rho-Hsu4-S}
\end{figure}
\begin{figure}[htb!]
    \centering  \includegraphics[width=0.55\textwidth]{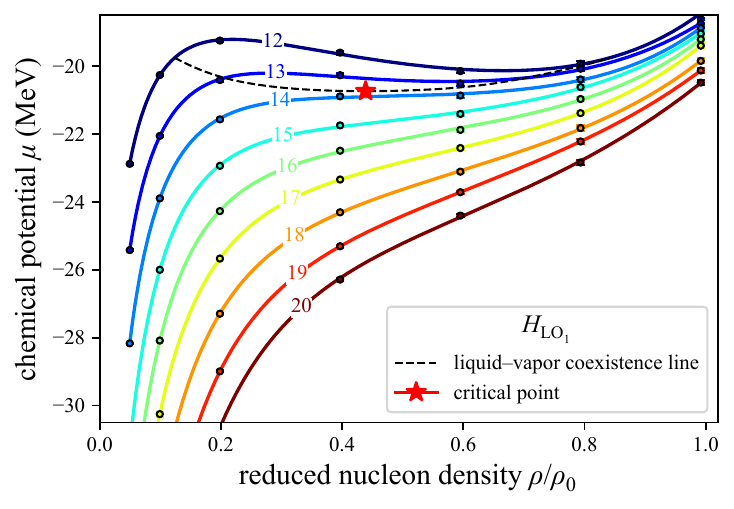}
\caption{Same as in Fig.~\ref{fig:mu-vs-rho-Hsu4}, but for the ``LO pionless-EFT'' Hamiltonian $H_{\mathrm{LO}_{1}}$.}
\label{fig:mu-vs-rho-HLO-1}
\end{figure}
\begin{figure}[htb!]
    \centering  \includegraphics[width=0.55\textwidth]{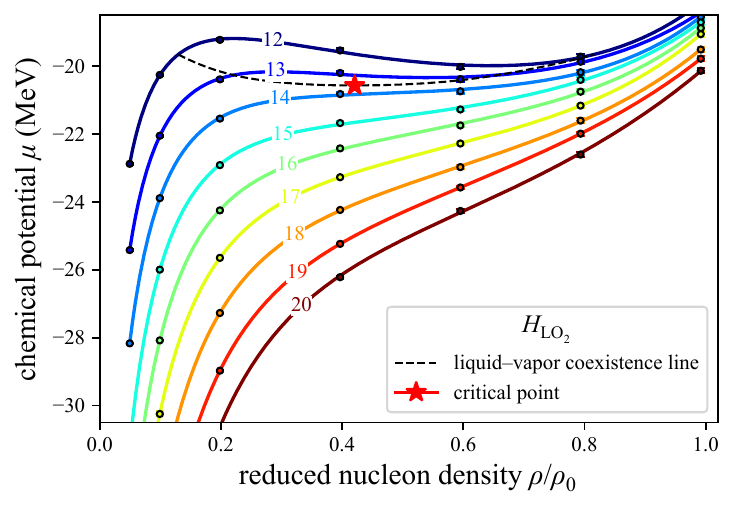}
\caption{Same as in Fig.~\ref{fig:mu-vs-rho-Hsu4}, but for the ``LO pionless-EFT'' Hamiltonian $H_{\mathrm{LO}_{2}}$.}
\label{fig:mu-vs-rho-HLO-2}
\end{figure}
\begin{figure}[htb!]
    \centering  \includegraphics[width=0.55\textwidth]{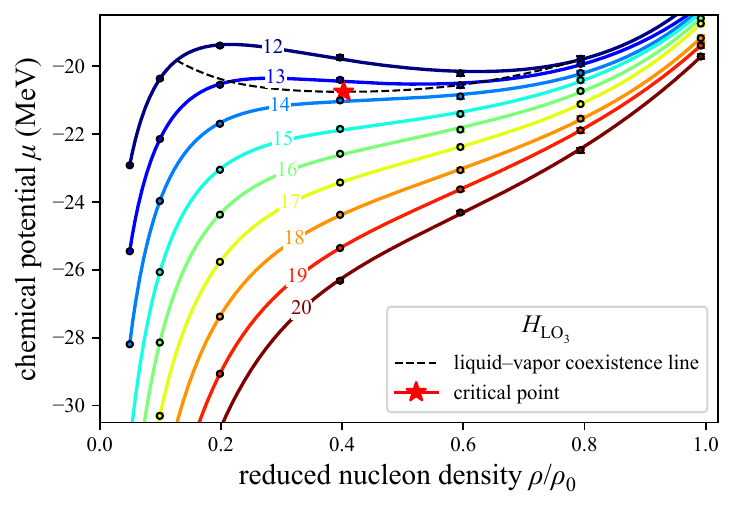}
\caption{Same as in Fig.~\ref{fig:mu-vs-rho-Hsu4}, but for the ``LO pionless-EFT'' Hamiltonian $H_{\mathrm{LO}_{3}}$.}
\label{fig:mu-vs-rho-HLO-3}
\end{figure}
Figures~\ref{fig:mu-vs-rho-Hsu4}, \ref{fig:mu-vs-rho-Hsu4-S}, \ref{fig:mu-vs-rho-HLO-1}, \ref{fig:mu-vs-rho-HLO-2}, and \ref{fig:mu-vs-rho-HLO-3} show the chemical potential as a function of the reduced density $\rho/\rho_{0}$ for the different Hamiltonians considered in this work, where $\rho_{0}=0.164$~fm${}^{-3}$. The symbols are the lattice results at discrete temperatures, while the solid curves are obtained from the fifth-order virial-like fits of the individual isotherms. The dashed curve denotes the liquid-vapor coexistence line extracted from the Maxwell construction, and the star marks the corresponding critical point.

The benchmark analysis presented in Sec.~\ref{sec:Benchmark} shows that the first-order perturbative pinhole-trace treatment remains quantitatively under control in the temperature and density regime relevant for the present work. The differences observed among the physical Hamiltonians should therefore be interpreted as genuine interaction effects rather than artifacts of the perturbative procedure. Taken together, our results show that as the interaction is refined from $H_{\mathrm{SU(4)}}$ to the LO Hamiltonians, the zero-temperature description improves, both in finite nuclei and in symmetric nuclear matter, while the critical temperature shifts downward. Within the present framework, the separation between zero-temperature improvement and finite-temperature criticality is systematic. In other words, the LO Hamiltonians improve finite-nucleus binding energies and move the saturation point toward the empirical region while reducing $T_{c}$ by about $1.6$--$1.8$~MeV relative to $H_{\mathrm{SU(4)}}$. The comparatively closer agreement of $H_{\mathrm{SU(4)}}$ with the empirical estimate of the critical temperature may therefore be accidental within this Hamiltonian family, and the present results suggest that further refinements will be needed to better describe the phenomenological critical region.

Our finite-temperature results can be compared with several earlier microscopic calculations. In Ref.~\cite{Baldo:1999cvh}, finite-temperature EoS calculations in the Bloch-De Dominicis/Brueckner-Hartree-Fock (BHF) framework with the Argonne $v_{14}$ interaction yielded $T_{c}\approx21$~MeV, with only a small reduction to $T_{c}\approx20$~MeV when a phenomenological three-body force was included. By contrast, thermodynamically consistent self-consistent Green's-function (SCGF) calculations with CD-Bonn or Nijmegen interactions plus an averaged Urbana three-body force found a much stronger reduction, from $T_{c}=18$--$20.5$~MeV with $2N$ forces alone to $T_{c}=12.5$--$11.5$~MeV once $3N$ forces were included~\cite{Soma:2009pf}. Ref.~\cite{Carbone:2018kji} studied five chiral $2N$+$3N$ Hamiltonians in both SCGF and BHF, and found that Hamiltonian dependence dominates many-body uncertainty, quoting an overall estimate $T_{c}=16.4\pm2.3$~MeV. Ref.~\cite{Keller:2022crb} reported $T_{c}=15.9$--$16.3$~MeV in finite-temperature chiral many-body perturbation theory (MBPT). Our lattice results therefore lie within the broader spread of microscopic predictions and show, within this lattice Hamiltonian family, that lower $T_{c}$ accompanies an improved zero-temperature description relative to $H_{\mathrm{SU(4)}}$, rather than being fixed by saturation and binding alone.

\section{Summary\label{sec:summary}}

The main question addressed in this work is whether the liquid-gas critical point of symmetric nuclear matter follows the same interaction trends as standard zero-temperature benchmarks when the sign-friendly lattice Hamiltonian is refined. To answer this question, we combined the pinhole-trace algorithm with a first-order perturbative treatment and studied a sequence of Hamiltonians ranging from $H_{\rm SU(4)}$ to Hamiltonians with physical  ${}^{1}S_{0}$ and ${}^{3}S_{1}$ channel dependence and three improved leading-order interactions. Benchmark calculations for representative Hamiltonian splittings show that the perturbative treatment is quantitatively reliable in the density and temperature regime relevant to the present analysis.

Using this framework, we extracted the finite-temperature EoS and the liquid-gas critical point of symmetric nuclear matter, and we compared these results with the zero-temperature saturation point of symmetric nuclear matter and the binding energies of selected nuclei. Within the present family of sign-friendly lattice Hamiltonians, the SU(4)-symmetric interaction yields the highest critical temperature, $T_c = 15.33(6)$ MeV. The inclusion of the ${}^{1}S_{0}$ and ${}^{3}S_{1}$ terms lowers it to $T_c = 15.13(18)$ MeV, while the three improved leading-order Hamiltonians give $T_c = 13.50(17)$--$13.71(19)$ MeV. At the same time, these LO Hamiltonians improve the overall description of finite-nucleus binding energies and move the zero-temperature saturation point toward the empirical region. The empirical critical-point values used for comparison should, however, be viewed as phenomenological reference estimates extracted indirectly from finite, dynamical, Coulomb-affected systems. Our analysis therefore shows that improving the ground-state sector does not automatically improve the finite-temperature critical region, and that liquid-gas criticality provides a complementary thermodynamic benchmark for future lattice interaction design.

The natural next steps are to extend the finite-temperature calculation to higher-order and higher-fidelity lattice interactions, to examine higher-order perturbative corrections where feasible, and to quantify the remaining theoretical uncertainties more systematically. It will also be important to determine whether the interaction trends found here persist in asymmetric matter and in observables that probe clustering and critical behavior more directly. These directions would sharpen the role of finite-temperature criticality as a benchmark for future lattice interaction design.

\acknowledgements
We thank members of the Nuclear Lattice Effective Field Theory Collaboration for useful discussions. We are grateful to Ulf-G. Meißner for useful comments on the manuscript. This work was supported by the Scientific and Technological Research Council of Turkey (TÜBİTAK) under Project No.~123F464. Computational resources were provided by TÜBİTAK ULAKBİM High Performance and Grid Computing Center (TRUBA). The numerical calculations reported in this paper were partially performed using the EuroHPC Joint Undertaking (EuroHPC JU) supercomputer MareNostrum~5, hosted by the Barcelona Supercomputing Center (BSC). Access to MareNostrum 5 was provided through a national access call coordinated by TÜBİTAK. We gratefully
acknowledge BSC, TÜBİTAK, and the EuroHPC JU for providing access to these resources and
supporting this research.

\appendix

\section{Euclidean-time extrapolation for finite nuclei and symmetric nuclear matter}

\begin{figure}[htb!]
    \centering  \includegraphics[width=0.9\textwidth]{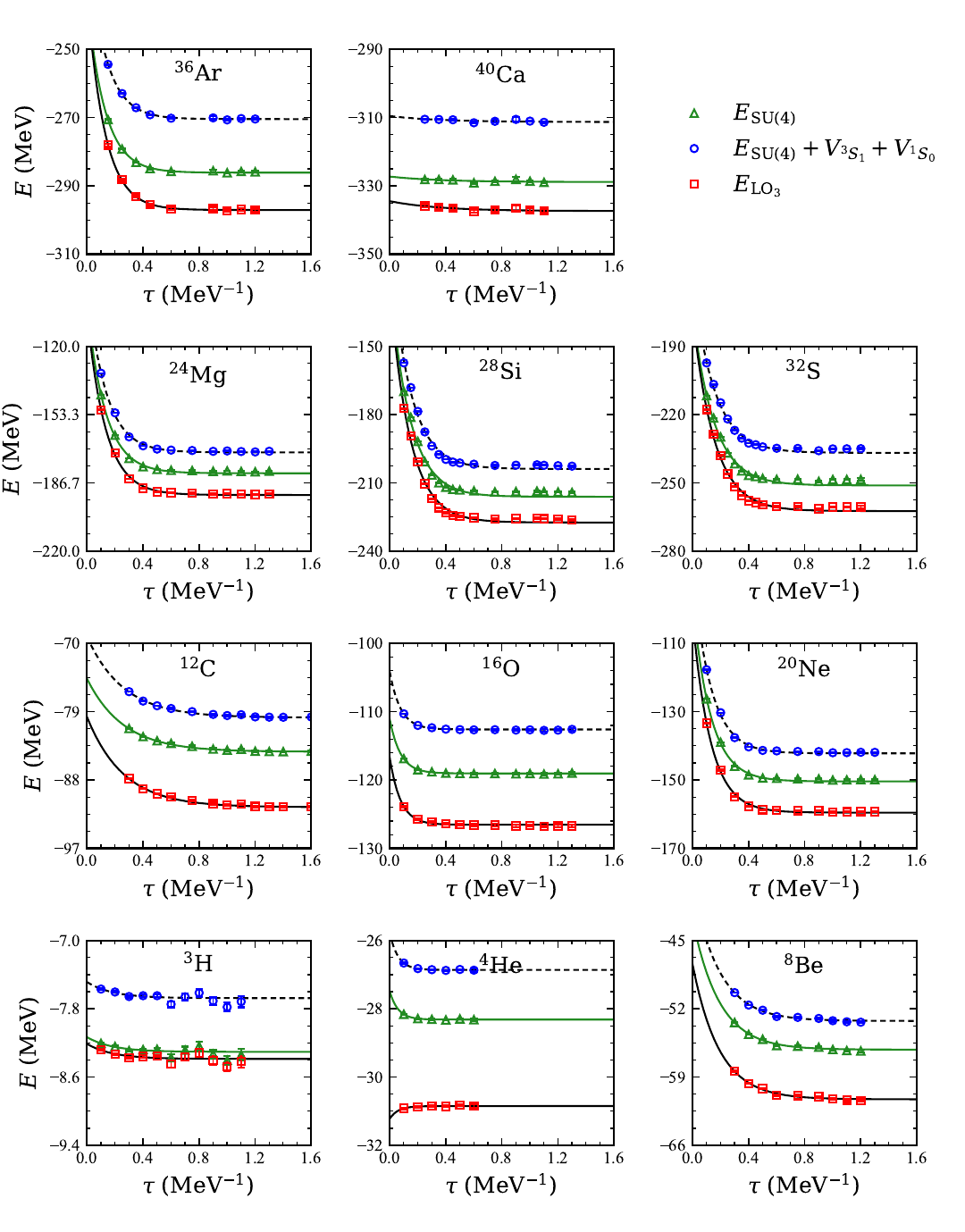}
\caption{Euclidean-time extrapolations of the finite-nucleus ground-state energies used in the present work. For each nucleus, the symbols show the lattice data for the different Hamiltonian choices, while the solid curves denote the corresponding fits used to extract the asymptotic energies.}
\label{fig:appendix-nuclei-euclid}
\end{figure}

\begin{figure}[htb!]
    \centering  \includegraphics[width=1.00\textwidth]{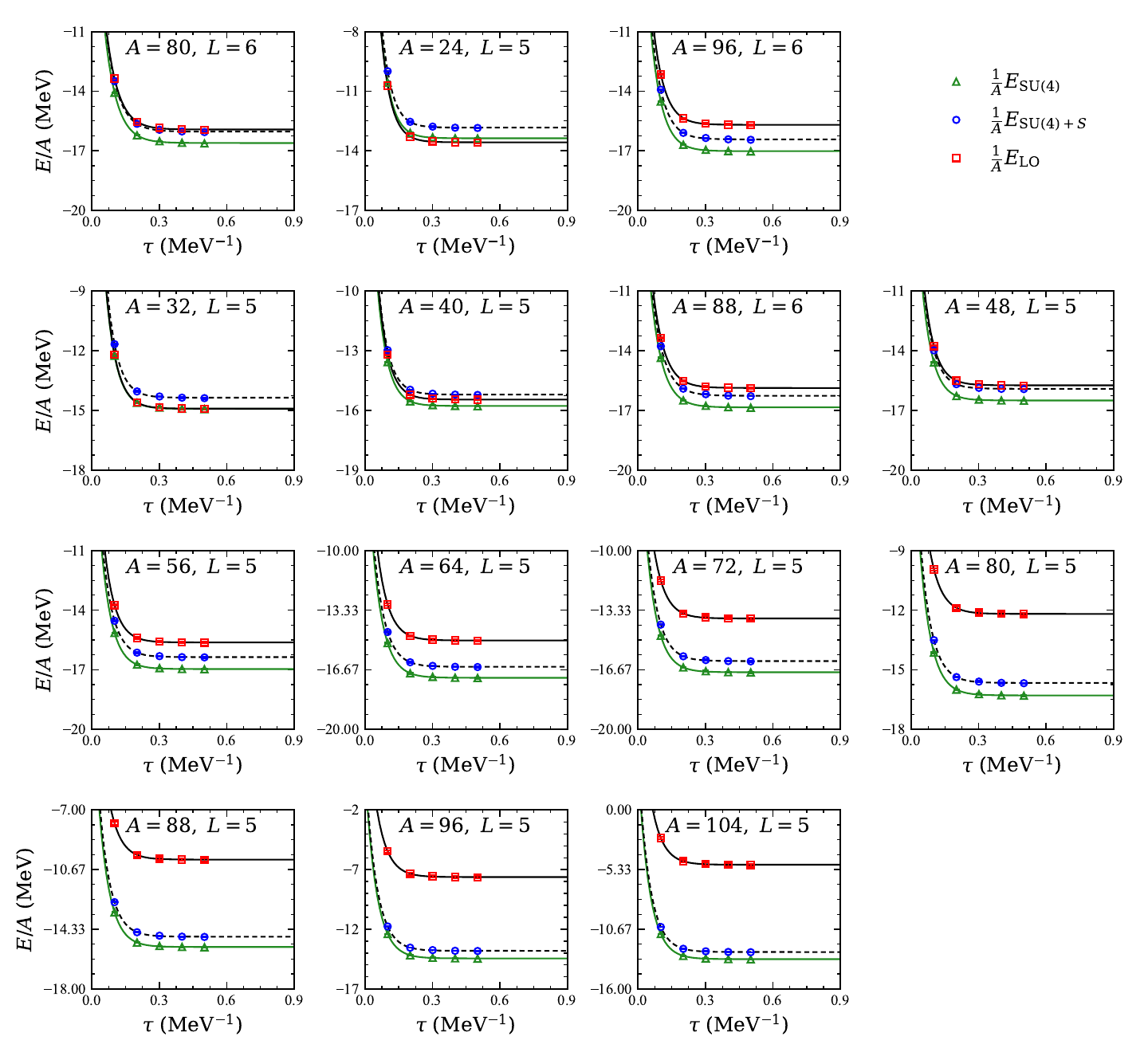}
\caption{Euclidean-time extrapolations of the symmetric nuclear matter energies per nucleon for the systems and box sizes considered in this work. The symbols show the lattice results, and the solid curves denote the fitted extrapolations used to obtain the asymptotic energies.}
\label{fig:appendix-snm-euclid}
\end{figure}

In this appendix we summarize the Euclidean-time extrapolations used to extract the zero-temperature energies reported in the main text. Fig.~\ref{fig:appendix-nuclei-euclid} shows the extrapolations for the selected finite nuclei, and Fig.~\ref{fig:appendix-snm-euclid} shows the corresponding extrapolations of the symmetric nuclear matter energies per nucleon for the particle numbers and box sizes considered in this work. In each case, the symbols denote the lattice Monte Carlo results at finite Euclidean time, and the solid curves are the fits used to determine the asymptotic large-$\tau$ values quoted in Tabs.~\ref{tab:critical_saturation_all} and \ref{tab:binding_energies_samples}. These extrapolations provide the zero-temperature input for the comparisons of binding energies and saturation properties discussed in Sec.~\ref{sec:results-discussions}. 

\bibliography{References}
\end{document}